\documentclass[aps,prx,superscriptaddress,floatfix,twocolumn,10pt]{revtex4-2}
\usepackage{graphicx}
\usepackage{physics}
\usepackage[normalem]{ulem}
\usepackage[utf8]{inputenc}
\usepackage{amsfonts}
\usepackage{xcolor}
\usepackage{hyperref}

\newcommand{\heff}{\hbar_\text{eff}}
\newcommand{\Pp}{\mathcal{P}}
\newcommand{\T}{\mathcal{T}}

\begin{document}

\title{
Degeneracy beyond the parity-symmetry protection in one-dimensional spinless models: The parity-violating Kerr parametric oscillator
}

\author{Jamil Khalouf-Rivera}
\affiliation{Departamento de Física, Facultad de Ciencias, Universidad de Córdoba, Campus de Rabanales, Edif. Einstein (C2), Córdoba, 14071, Spain}
\affiliation{Depto. de Ciencias Integradas y
  Centro de Estudios Avanzados en Física, Matemáticas y Computación, Universidad de Huelva, Huelva
  21071, SPAIN}
\author{Miguel Carvajal}
\author{Francisco Pérez-Bernal}
\affiliation{Depto. de Ciencias Integradas y
  Centro de Estudios Avanzados en Física, Matemáticas y Computación, GIFMAN, UHU, Unidad Asociada al CSIC por el IEM, Universidad de Huelva, Huelva
  21071, SPAIN}
\affiliation{Instituto Carlos I de Física Teórica y Computacional,
  Universidad de Granada, Granada 18071, SPAIN}

\begin{abstract} 
    One-dimensional quantum systems that undergo spontaneous symmetry-breaking, having a symmetric (non-degenerate) and 
 a broken-symmetry (doubly-degenerate) phase, have been intensely studied in different branches of physics. In most cases, the spontaneously-broken symmetry is parity. However, it is possible to obtain similar phases in systems without parity symmetry, through an antiunitary symmetry that implies a two-fold symmetry either on momentum or coordinate in the system's classical limit. To illustrate this phenomenon, we use a Kerr parametric oscillator (KPO) with one- and two-photon drives that, despite the breaking of parity symmetry, may have doubly-degenerate levels. Different realizations of squeezed  KPOs convey a great deal of attention, as effective Hamiltonians for driven superconducting circuits and the occurrence of degeneracy in such systems could be of practical interest in their application to obtain protected qubits in parity-breaking setups. In addition to this, the reported spectral features strongly indicate the existence of additional symmetries in the system. 
\end{abstract}
\maketitle

\paragraph{Introduction to the problem.---} Symmetries are a fundamental guiding principle in physics and chemistry and they have multiple facets, e.g.,~geometrical, permutational, dynamical symmetries or supersymmetries~\cite{wigner2013group,Hamermesh1962,SolMolPhys1974, iachello2011,Schwichtenberg2015,SymInPhys}. In quantum systems, the von Neumann-Wigner's theorem~\cite{Neumann-Wigner} determines that, for unitary transformations, the Hilbert space states can be labeled with the irreducible representations (irreps) of the system's symmetry group. Hence, the  Hamiltonian matrix for such systems can be split into different blocks, corresponding to each of the symmetry group irreps~\cite{Bargmann1964}. As a consequence of this, degeneracy appears if the state resulting from the application of a symmetry group operation to a system's eigenstate is linearly independent from the original state. A clear example of this are parity-conserving one-dimensional (1D) systems, like the quartic oscillator, where symmetry can be spontaneously broken and degeneracy appears in the broken-symmetry phase \cite{ Gilmore1986, Milburn1997, Steel1998, Franzosi2001, Review_BJJ}. However, if the system symmetry does not allow a matrix representation, as in the anti-unitary symmetries case, the system Hamiltonian cannot be divided into smaller blocks, hindering the identification of  degenerate or quasi-degenerate subspaces.

It is interesting to consider that, in the case of potentials with symmetric wells separated by finite-height energy walls, the tunneling-transition probability can be very small, but it is never exactly zero. It is well known, from the WKB approach, that the energy gap between such states decreases exponentially with the barrier height~\cite{Landau1991}. Therefore,  the degeneracy of states belonging to different symmetries in the broken-symmetry phase is not an exact degeneracy and such states are often dubbed as quasi-degenerate~\cite{Un_degeneration}. The calculation of energy gaps in such systems requires arbitrary precision calculations to avoid running into arithmetic underflow in numerical calculations. In such cases, the degeneracy stems from the existence in the classical limit of two (or more) disconnected and symmetric trajectories in canonical coordinate-momentum space.

\begin{figure*}
    \centering
    \includegraphics[width=\textwidth]{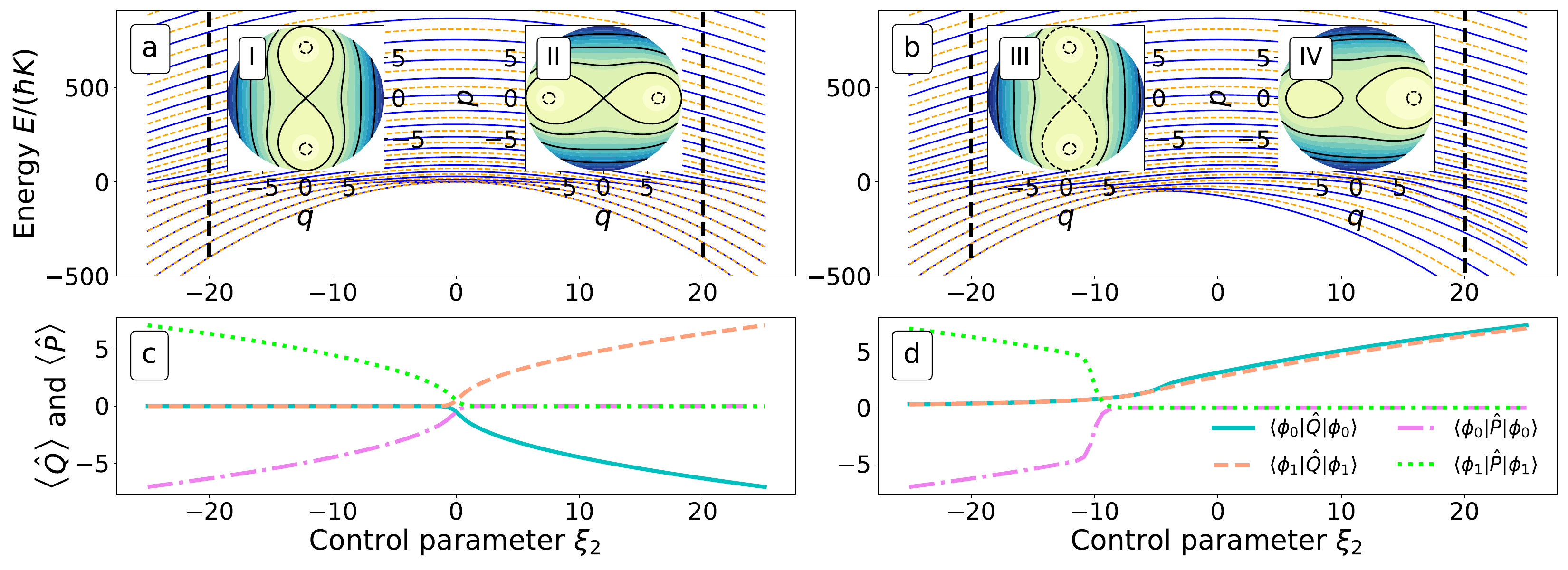}
    \caption{Panels (a) and (b) correspond to the truncated spectra of Hamiltonian \eqref{eq:kerrH} as a function of the  two-photon squeezing amplitude $\xi_2$  with $\xi_1=0$ (parity-conserving case) and $\xi_1=-30/\sqrt{2}$. 
    The inset panels (I-IV) correspond to the energy contours of Hamiltonian \eqref{eq:classH} for the cases marked by dashed vertical lines with parameters $(\xi_2,\xi_1)$: I (-20,0), II (20,0), III (-20,-30$/\sqrt{2}$), and IV (20,-30$/\sqrt{2}$). Panels (c) and (d)  show the expectation value of coordinate and momentum operators, $\hat Q$ and $\hat{P}$, for the two lowest energy eigenstates, $\ket{\phi_0}$ and $\ket{\phi_1}$, versus the control parameter $\xi_2$ for $\xi_1 = 0$ in panel (c) and $-30/\sqrt{2}$ in panel (d). Quantum calculations have been carried out with $N_e=1$.}
\label{fig:Correlationxi2}
\end{figure*}

\begin{figure}
    \centering
    \includegraphics[width=\columnwidth]{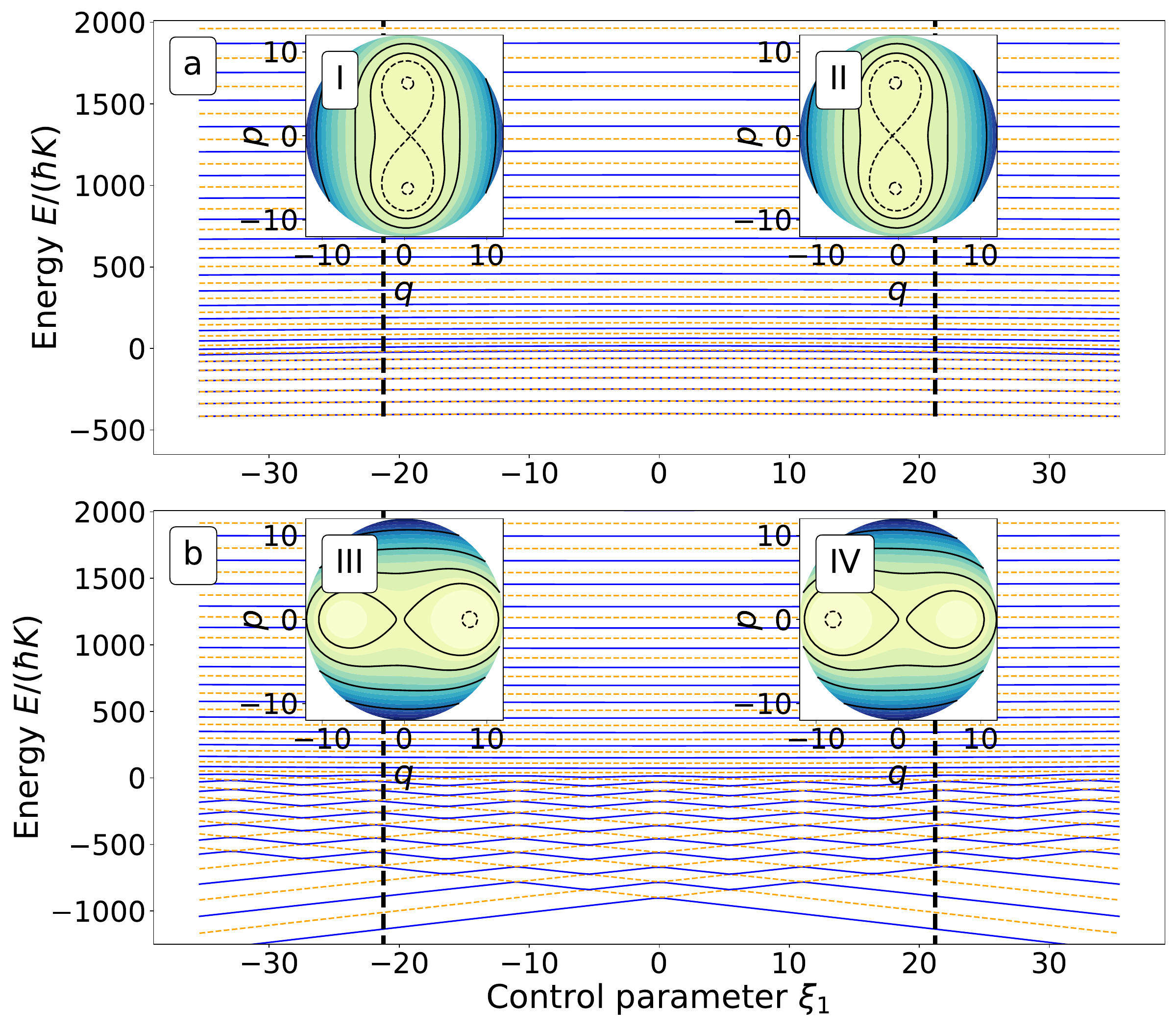}
    \caption{Panels (a) and (b) correspond to the truncated spectra of Hamiltonian \eqref{eq:kerrH} as a function of the  one-photon squeezing amplitude $\xi_1$  with $\xi_2=-30$ and $\xi_2=30$, respectively. 
    The inset panels (I-IV) contain the energy contours of Hamiltonian \eqref{eq:classH} for the cases marked by dashed vertical lines in panels (a) and (b), with parameters $(\xi_2,\xi_1)$: I ($-30$, $-30/\sqrt{2}$), II ($-30$, $30/\sqrt{2}$), III ($30$, $-30/\sqrt{2}$), and IV ($30$, $30/\sqrt{2}$). Quantum calculations have been carried out with $N_e=1$.}
    \label{fig:Correlationxi1}
\end{figure}
As already mentioned, if we exclude accidental degeneracies, the occurrence of energy doublets in a system implies the existence of a symmetry that explains such quasi-degeneration. Usually such symmetry, in 1D systems, is parity, which implies the simultaneous sign reversal of the coordinate and its associated momentum, $\mathcal{P}:\left\{q,p\right\}\to\left\{-q,-p\right\}$~\cite{Bishop1994}. Our selected model of study is a non-linear Kerr parametric oscillator, a system that currently receives an intense heed due to its role as effective Hamiltonian operators for driven superconducting circuits of relevance in quantum computing and quantum simulation~\cite{Bartolo2016, Zhang2017, Puri2017, Goto2019, npj2023PB, Iachello2023, Minganti2023, Reynoso2023, GarciaMata2024, Iachello_2024, QWei2020, Reynoso2025} with experimental realizations~\cite{Sivak2019, Grimm2020, Blais2021, Jaya2022, Cortinas2023Pi, beaulieu2023, Marti2024, Iyama2024, Dutta2024, Frattini2024,Albornoz2024, Masuda2025}. We have included one- and two-photon squeezing terms in the model Hamiltonian to break up parity symmetry~\cite{Bartolo2016,QWei2020}, a setup that can be experimentally achieved \cite{Albornoz2024}.
In broad lines, we show how a degeneracy that scales exponentially with the system size can be observed in a 1D quantum system that does not conserve parity symmetry. This fact can have important  consequences for quantum calculations, where the adiabatic approximation is considered, or for the protection of quantum states.

The present letter is organized as follows. First, we introduce the chosen model to illustrate our results; then, we study the energy spectra, their classical limit, and the expectation value of coordinate and momentum operators for selected states, comparing the results obtained in the parity-symmetric and parity-broken cases. Finally, we study the scaling of the broken-symmetry phase energy doublets as we drive the system towards its classical limit, showing that energy gaps scale identically in systems with and without parity symmetry.

\paragraph{The squeeze-driven non-linear Kerr oscillator.---}
Double-well quantum systems have been experimentally accessible using {the well-known Josephson effect\cite{Review_BJJ,Carapella2004} and  driven superconducting circuits~\cite{Mooij1999,Rouse1995,Caspar2000}. A possible experimental realization of a parity-symmetric double-well system can be achieved using a microwave-driven superconducting circuit that, in the rotating wave approximation, has an effective, time-independent, Kerr Hamiltonian with a two-photon squeezing term~\cite{Grimm2020,beaulieu2023,Marti2024,Iyama2024,Frattini2024}. Parity symmetry can be broken by adding a one-photon drive term, obtaining a model for non-symmetric double wells~\cite{Albornoz2024}. 

The Hamiltonian for a non-linear Kerr parametric oscillator including one- and two-photon drives is~\cite{Bartolo2016,QWei2020,Albornoz2024}
\begin{equation}\label{eq:kerrH}
    \frac{\hat{H}(\xi_2,\xi_1)}{K\hbar}=  \frac{\hat{a}^{\dagger 2}\hat{a}^2}{N_e}- \xi_2\left(\hat{a}^{\dagger 2} + \hat{a}^2\right) +  \xi_1\sqrt{N_e}\left(\hat{a}^\dagger + \hat{a}\right)~,
\end{equation}
where $\hat{a}^\dagger$ and $\hat{a}$ are boson creation and annihilation operators, $K$ is the Kerr nonlinearity, $\xi_1$ and $\xi_2$ are dimensionless control parameters associated with the one- and two-photon drives, and $N_e$ is a parameter that, for large values, drives the system to its classical limit. The global energy scale in Hamiltonian~\eqref{eq:kerrH} varies linearly with $N_e$. The coordinate and momentum operators associated with the creation and annihilation operators are $\hat{Q}=\sqrt{1/(2N_e)}\left(\hat{a}^\dagger + \hat{a}\right)$ and  $\hat{P}=i\sqrt{1/(2N_e)}\left(\hat{a}^\dagger - \hat{a}\right)$. 
If the one-photon drive control parameter $\xi_1$ is zero, Hamiltonian~\eqref{eq:kerrH} is parity-invariant, $\hat{\Pp} \hat{H}(\xi_2,\xi_1=0)\hat{\Pp}^{-1}=\hat{H}(\xi_2,\xi_1=0)$, and the parity operator can be expressed as $\hat{\Pp}=e^{i \pi \hat{n}}=e^{i\pi\hat{a}^\dagger\hat{a}}$. Therefore, if we solve the problem using a Fock basis, $\ket{n}\propto \left(\hat{a}^\dagger\right)^n \ket{0}$, the system Hamiltonian can be split into two non-mixing blocks, for even and odd parity states~\cite{Chavez2023, Iachello2023}.  Such symmetry is broken by the one-photon drive term, proportional to the coordinate $\hat{Q}$.

The classical limit of the chosen model can be obtained using the transformation $a\to\sqrt{N_e/2}\left(\hat{Q}+i \hat{P}\right)$. Notice that  $[\hat{Q}, \hat{P}] = i/N_e$ and both quantities commute in the large $N_e$ limit. The classical limit of Hamiltonian \eqref{eq:kerrH} is obtained as $h_\text{class}(q,p)=\lim_{N_e\to\infty}\frac{H}{N_eK\hbar}$,
\begin{equation}\label{eq:classH}
    h_\text{class}(q,p)=\frac{1}{4}\left(q^2+p^2\right)^2 - \xi_{2} \left(q^2-p^2\right)+\xi_{1} \sqrt{2}q~.
\end{equation}

For $\xi_1$ values different from zero, the parity symmetry is broken and the classical Hamiltonian in Eq.~\eqref{eq:classH} is only invariant under a reflection on the $q$ axis, $\T:\left(q,p\right)\to\left(q,-p\right)$, which is associated with a time-reversal operator $\hat{\T}$, a symmetry operation that explains the appearance of doubly-degenerate states (Kramer's doublets) in systems with a half-integer total spin \cite{messiah2020,Kramers1930,Klein1952,Rosch1983}. As it can be easily demonstrated, in the case of systems with integer total spin, the application of the time reversal operator to a system's eigenstate does not produce an independent state, and no degeneracy is possible. This also happens in spinless potential systems with a single degree of freedom and a classical limit Hamiltonian $h(q,p) = p^2/2m + V(q)$. However, in systems with a more complex dependence on the momentum, symmetries in the full phase space should be taken into consideration, as it has already been explored in phase space crystals~\cite{Guo2013, Guo2022, Hannaford2022,Kopaei2023}. In our case, the anharmonic nature of the Kerr parametric resonator Hamiltonian results in  a momentum dependence that allows for the existence of quasi-degenerate states associated with the time-reversal symmetry. 
The parity transformation $\Pp$ can be considered as the product of a reflection  on the $q$ axis, $\T$, and a reflection on the $p$ axis, $\Pp\T:(q,p)\to(-q,p)$. The coordinate and position commutator ought to be conserved under any Hamiltonian symmetry operation. If $\hat{\mathcal{O}}=\hat{\T}$ or $\hat{\Pp}\hat{\T}$, $[\hat{ \mathcal{O}}\hat{Q} \hat{\mathcal{O}}^{-1},\hat{\mathcal{O}}\hat{P}\hat{\mathcal{O}}^{-1}]=-[\hat{Q}, \hat{P}]=\hat{\mathcal{O}}(i/N_e)\hat{\mathcal{O}}^{-1}=-i/N_e$. Therefore, both $\hat{\T}$ and $\hat{\Pp}\hat{\T}$ are antiunitary and antilinear symmetry operations that, when acting on a complex constant $c\in\mathbb{C}$, results in its complex conjugate $\mathcal{O} c={c^*}\mathcal{O}$~\cite{messiah2020}.

Hence, for nonzero $\xi_1$ values, parity is broken, but Hamiltonian \eqref{eq:kerrH} still conserves the time-reversal symmetry, $\hat{\T} \hat{H}(\xi_2,\xi_1)\hat{\T}^{-1}=\hat{H}(\xi_2,\xi_1)$. If the single-photon drive operator is proportional to $\hat{P}$ instead of $\hat{Q}$, the conserved symmetry is $\hat{\Pp}\hat{\T}$ and the obtained results are equivalent to the present ones.

\begin{figure}
    \centering
    \includegraphics[width=\linewidth]{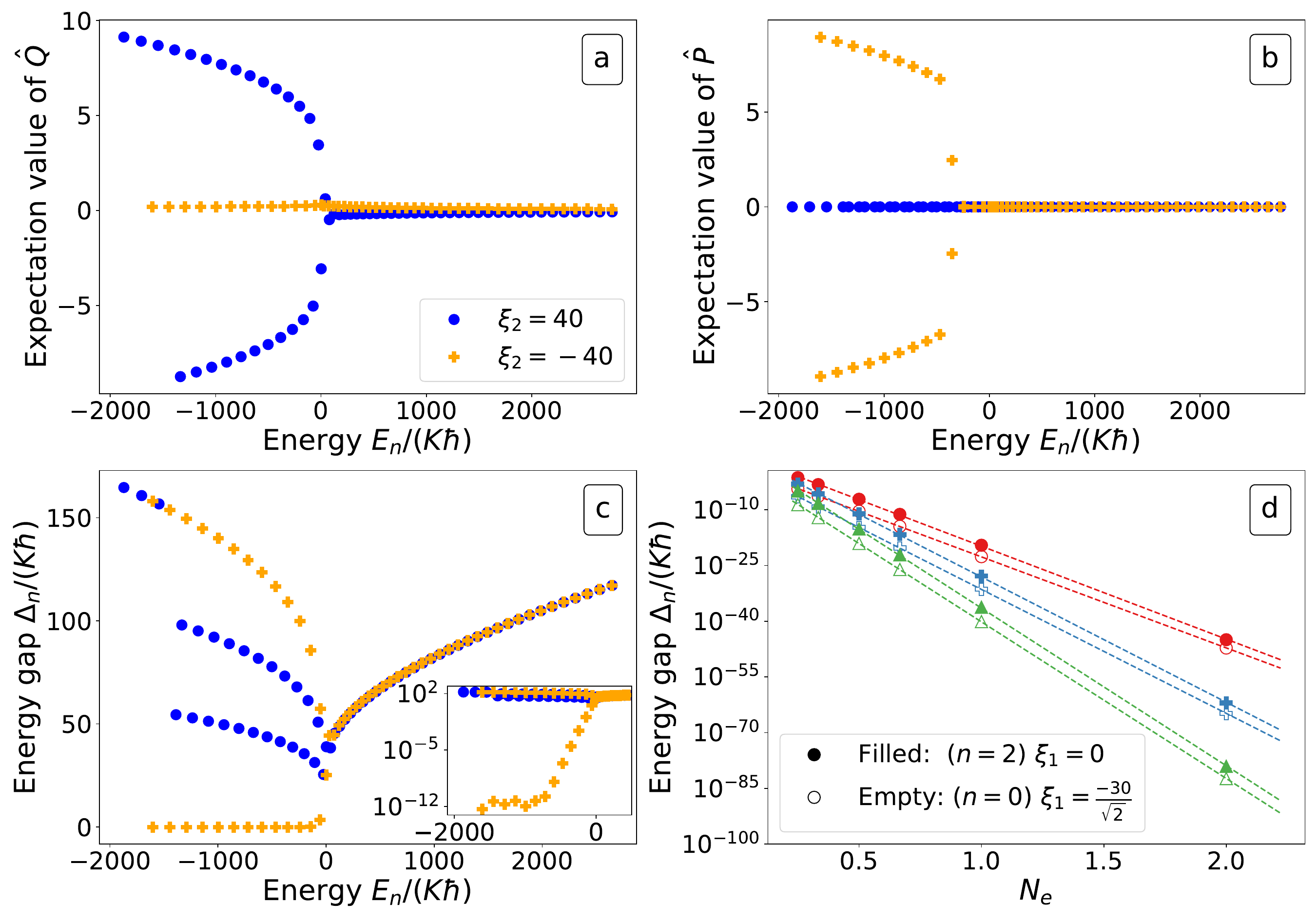}
    \caption{Panels (a) and (b) depict the expectation value of $\hat Q$ and $\hat P$, while panel (c) shows the energy gap of adjacent levels, $\Delta_j = E_{j+1} - E_j$, versus the system energy for $(\xi_2=40,\xi_1=-30/\sqrt{2})$ (blue points) and $(\xi_2=-40,\xi_1=-30/\sqrt{2})$ (orange crosses), and $N_e=1$. 
    The inset in panel (c) is a zoom of the main plot using logarithmic scale. In panel (d), we show the energy gap $\Delta_2$ of different parity-symmetric systems ( $\xi_1 =0$, filled markers) and $\Delta_0$ of different parity-deformed systems ($\xi_1=-30/\sqrt{2}$, empty markers) using log-lin scales.  The values of $\xi_2$ are $-30$ (red circles), $-40$ (blue crosses), and $-50$ (green triangles). The dashed lines correspond to the exponential fitting (see the main text).}
    \label{fig:system}
\end{figure}

 \paragraph{Results.---} In Fig.~\ref{fig:Correlationxi2}(a) we show the truncated energy spectrum of Hamiltonian~\eqref{eq:kerrH}  as a function of $\xi_2$ for $\xi_1 = 0$, using blue solid (orange dashed)  lines for even (odd) parity states. We can distinguish a high level-density separatrix line at a critical energy equals to zero, that marks the existence of an excited-state quantum phase transition, separating the spectrum into two different quantum phases~\cite{QWei2020, Chavez2023,Iachello2023,Reynoso2025}. The density of states at the critical energy diverges logarithmically in the mean-field limit of the system, see Ref.~\cite{esqpt_review} for a complete review on such transitions. Below the critical energy, states are quasi-degenerate, while these doublets get split above the critical energy. In panels Fig.~\ref{fig:Correlationxi2}(I) and (II) we plot the classical energy surfaces given by the Hamiltonian~\eqref{eq:classH} for $\xi_1 = 0$ and $\xi_2=-20$ (I) and $20$ (II). In both cases the system is parity invariant and has two symmetric minima that, for $\xi_2<0$ ($\xi_2>0$), lie along the $q=0$ ($p=0$) axis. Black lines are constant energy contours, i.e. phase-space trajectories accessible to the classical system for a fixed energy value. For energies below the critical energy, $E_c=0$, there are two disconnected trajectories invariant under the parity transformation $\Pp$, in accordance with Refs.~\cite{Chavez2023, Iachello2023}. Nonzero $\xi_1$ values break the parity symmetry with a resulting spectrum shown in Fig.~\ref{fig:Correlationxi2}(b), where we depict energy values as a function of $\xi_2$ for a one-photon drive parameter $\xi_1=-30/\sqrt{2}$. In this case, to facilitate the understanding of the figure, we keep using alternated blue solid and orange dashed lines for adjacent energy levels, but in this case the matrix Hamiltonian cannot be broken into blocks for even and odd states as in the previous case. It is clear from this figure how quasi-degenerate doublets appear in the lower end of the spectrum for negative values of the control parameter $\xi_2$. As in the parity-symmetric case, in insets Fig.~\ref{fig:Correlationxi2}(III) and (IV) we depict the classical energy surfaces for  $\xi_1 = -30/\sqrt{2}$ and $\xi_2=-20$ (III) and $20$ (IV). In both cases, the system conserves the $\T:(q,p)\to(q,-p)$ symmetry. When $\xi_2<0$, the two minima in 
Fig.~\ref{fig:Correlationxi2} (III) are equivalent and, for negative energies, there are two symmetric energy contours accessible to the system. In other words, the existence of two equivalent closed trajectories that are invariant under the symmetry transformation $\T$ explains the existence of {\em spectral kissing} in the parity-violating system. As previously mentioned, and unlike in the parity-symmetric  case, we cannot use the $\T$ symmetry irreps to label the resulting eigenstates~\cite{SolMolPhys1974,Rosch1983,wigner2013group,Geru2018TimeReversal}. 
To grasp a better understanding of the nature of the degeneration in the case with broken parity, we present the expectation value of $\hat{Q}$ and $\hat{P}$ for the two eigenstates with lowest energy, $\ket{\phi_i}$ with $i=0, 1$ in the different cases. In the parity-symmetric case, with $\xi_1=0$, once the Hamiltonian is split into even and odd symmetry blocks, the expectation value of both operators will be zero, as expected for even and odd wavefunctions. To avoid this, we can either combine the resulting quasi-degenerate eigenstates to locate the wave function in one of the two minima, or make them distinguishable adding a tiny interaction term $\epsilon\left(\hat{Q} + \hat{P}\right)$ to the Hamiltonian with a small enough value of $\epsilon$ \footnote{Notice that the added term should combine both operators. If we only added $\hat Q$ or $\hat P$, we would only break partially the parity symmetry and we would not obtain the desired results.}. The obtained results are depicted in Fig.~\ref{fig:Correlationxi2}(c). The expectation value of $\hat Q$ is zero for negative $\xi_2$ values and has symmetric positive and negative values for positive $\xi_2$ values. The results for $\hat{P}$ are reversed, zero for positive values of $\xi_2$, and symmetric positive and negative values for negative $\xi_2$ values. In Fig.~\ref{fig:Correlationxi2}(d), we show the expectation value of $\hat Q$ and $\hat P$ as a function of the control parameter $\xi_2$ for $\xi_1 = -30/\sqrt{2}$. The results in panel (c) have a stark contrast with the two non-symmetric curves observed in panel (d). However, the expectation value of $\hat P$ in panel (d) is still symmetric even for a nonzero $\xi_1$ value, as expected from the classical limit analysis of the $\T$-symmetric system.

To grasp a better understanding of the different situations we depict in Fig.~\ref{fig:Correlationxi1}(a) and (b) the truncated energy spectrum of Hamiltonian~\eqref{eq:kerrH}  as a function of $\xi_1$ for $\xi_2 = -30$ and  $\xi_2 = 30$, respectively, using once more blue solid (orange dashed)  lines for adjacent states. We include insets Fig.~\ref{fig:Correlationxi1}(I) to (IV) with the classical energy contours for selected $\xi_2$ values. It is clearly evinced the differences between the negative and positive $\xi_2$ cases. In Fig.~\ref{fig:Correlationxi1}(a) the low energy levels are quasi-degenerate and with a very smooth dependence on $\xi_1$ value. As seen in the corresponding energy contours, Fig.~\ref{fig:Correlationxi1}(I) and (II), the classical limit has two wells that are symmetric with respect to a reflection on the $p$ axis while the $\xi_1$ value breaks the symmetry with respect to the reflection on the $q$ axis. The situation, for the lower energy states, is completely different in the  $\xi_2 = 30$ case, as shown in  Fig.~\ref{fig:Correlationxi1}(b), with a more complex pattern of crossings. Beware of the difference between insets in  Figs.~\ref{fig:Correlationxi1}(a) and (b), the two trajectories with the same energy are symmetric in the first case and the symmetry is broken in the second case. Energy contours  Fig.~\ref{fig:Correlationxi1}(III) and (IV) reveal a phase space with two non-equivalent minima, that is consistent with the flow of energy levels in  Fig.~\ref{fig:Correlationxi1}(b), typical from a situation with two minima that reach coexistence at a critical value, $\xi_1 = 0$ in our case. Further details on the dependence of the spectrum of Hamiltonian~\eqref{eq:kerrH} with respect to the $\xi_1$ control parameter can be found in App.~\ref{app:QP}.

Once the quantum and classical limits are considered as a function of $\xi_1$ and $\xi_2$, we proceed to an in-depth analysis of the $\T$-symmetric system ($\xi_1=-30/\sqrt{2}$) for different values of the control parameter $\xi_2$. In Fig.~\ref{fig:system}(a), we display the expectation value of the observable $\hat{Q}$ versus the state energy for eigenstates of Hamiltonian \eqref{eq:kerrH} with $\xi_1=-30/\sqrt{2}$ and two $\xi_2$ values: $\xi_2=-40$ (orange crosses) and $\xi_2=40$ (blue points). In the $\xi_2=-40$ case, the expectation value only exhibits one branch that remains close to zero.  However, the  $\xi_2=40$ case exhibits two asymmetric branches for negative energy values, and a single line close to zero after such energy value. The situation varies if we observe $\hat{P}$ in panel (b) of Fig.~\ref{fig:system}. The $\T$-symmetric system exhibits two symmetric branches before the critical energy, and becomes zero above such critical energy, whereas the $\xi_2=40$ case remains at zero. This is easily understood considering that the two cases under consideration have a classical limit similar to the energy contours in Figs.~\ref{fig:Correlationxi1} (I) and (III). In Fig.~\ref{fig:system} (c) we show the energy gap of adjacent levels $\Delta_j=E_{j+1}-E_{j}$ for both systems.  For $\xi_2=-40$ (orange crosses), the system has  $\T$-reversal energy doublets for negative energy values and up to the critical energy and the anharmonicity turns into positive after the transition. In the $\xi_2=40$ case (blue points), we can explain the two curves before the critical energy because there are two asymmetric minima with bound-states. We have added an inset to this panel, using logarithmic scale in the ordinate axis, to highlight that the reported degeneracy is consistent with a standard double-precision floating-point calculation as ours.

In the KPOs case, of major technical relevance, the \textit{level kissing} concept was recently introduced~\cite{Frattini2024} with connections to  these quasi-degenerate energy doublets~\cite{Chavez2023, Iachello2023}. An exponential approach means that, for levels in the symmetry-broken phase, the energy gap tends exponentially to zero as the system approaches its classical limit, i.e.~$\Delta_j\propto e^{-\delta N_e}$. The numerical verification of this dependence is hindered by the high numerical precision requested and the need of exploring the convergence of the obtained results with the basis dimension. In Fig.~\ref{fig:system}(d), we depict the adjacent levels energy gap $\Delta_j$, in logarithmic scale, versus the $N_e$ parameter that drives the system to its classical limit for parity-symmetric systems with $\xi_1=0$ (solid markers) and parity-deformed systems with $\xi_1=-30/\sqrt{2}$ (empty markers), setting $\xi_2=-30$ (red circles), $\xi_2=-40$ (blue crosses), and $\xi_2=-50$ (green triangles). In the symmetric case, we have chosen the gap $\Delta_2=E_3-E_2$ (first excited state if we consider parity blocks) because the first two levels are exactly degenerate \cite{Puri2017}. Calculations in this case have been performed using arbitrary precision in Mathematica~\cite{Mathematica}.
In each case, we have checked the convergence of the depicted energy differences increasing simultaneously the precision of the calculation and the truncated Hilbert space dimension.  
 It is important to emphasize that the global energy scale increases as $N_e$ grows. The results of a fit to an exponential law,  are shown using dashed lines of the same colors than the points. The numerically obtained exponents are  shown in the third and fourth columns of Tab.~\ref{tab:deltas}, together with an analytical estimation. Such estimation, that approximates $\delta$ to $\delta_\text{app}=2\left|\xi_2\right|$, can be obtained by studying the overlap in the parity-symmetric system of two coherent states, locating one in each well (See App.~\ref{App:class} for a further description). 

\begin{table}
    \centering
    \begin{tabular}{|c|c|c|c|}
         \hline
         $\xi_2$ & $\delta_\text{app}(\xi_1=0)$ & $\delta(\xi_1=0)$ & $\delta(\xi_1=-30/\sqrt{2})$ \\
         \hline
         -30 & 60 & $57.6\pm 0.4$ &  $56.42\pm 0.09$ \\
         -40 & 80 & $77.6\pm 0.4$ & $76.81 \pm 0.10$ \\
         -50 & 100& $97.7\pm 0.4$ & $97.08 \pm 0.10$ \\
         \hline
    \end{tabular}
    \caption{Estimated $\delta_\text{app}=2\left|\xi_2\right|$, second column. Optimized $\delta$ parameters for parity-symmetric ($\xi_1=0$, third column) and parity-deformed ($\xi_1=-30/\sqrt{2}$, fourth column) systems.}
    \label{tab:deltas}
\end{table}

\paragraph{Conclusions.---}
The main conclusion of the present work is that quasi-degenerate energy doublets may happen in parity-breaking spinless 1D systems that have a complex enough dependence on the momentum in their classical limit. In such cases, these doublets appear in the broken-symmetry phase of systems with  $\T$ or $\Pp\T$ symmetries. We have checked that the energy gap on quasi-degenerate levels tends exponentially to zero as the system tends to the classical limit in the same way than in $\Pp$-symmetric systems.  

Apart from the evident interest of the existence of such degeneracy from a theoretical point of view, the present results can be of interest to a broader community. In particular, superconducting circuits are nowadays  one of the most promising experimental devices in the realization of quantum computers and quantum simulators \cite{Devoret2005, Devoret2013, Stassi2020, Huang2020, Blais2021, Bravyi2022, Copetudo2024, Dutta2024}. We have selected Hamiltonian \eqref{eq:kerrH} to illustrate our results because the combination of a Kerr resonator plus a two-photon squeezing term is an effective Hamiltonian associated to driven KPOs. Recent experimental results in the field of superconducting circuits, published by A. C. C. de Albornoz and collaborators ~\cite{Albornoz2024}, have shown that it is possible to obtain a quantum analog simulator for a system with asymmetric wells, with a classical limit equivalent to the phase space energy surface in Fig.~\ref{fig:Correlationxi2}, with a driven Kerr parametric oscillator including a third order non-linearity. The authors focused on the level crossing for the parity-violating system but from this reference it is clear that the study of {\em spectral kissing} protected by time reversal is beyond current experimental access.  The existence of such quasi-degenerate levels could provide an unexpected stabilization to quantum states in systems that do not conserve parity. 
The recent developments in QPTs for open KPOs~\cite{Bartolo2016,Iachello2024} indicate that the extension of the present work to open systems is also a promising research venue.

Finally, we also expect that our results will be also useful to researchers working in different quantum computing frameworks that rely on the quantum adiabatic theorem (e.g. quantum annealing and other). A careful consideration of all possible symmetries is of the utmost importance in such cases,  to be aware of possible problems with the adiabatic approximation. 

\begin{acknowledgments}
The authors thank Profs. Francesco Iachello and Lea F.~Santos for their valuable comments and suggestions. This project received funding through Grant No. PID2022-136228NB-C21 funded by
MICIU/AEI/10.13039/501100011033 and, as appropriate, by “ERDF A way of making Europe, by ERDF/EU,”
by the European Union, or by the European Union NextGenerationEU/PRTR. This research has also received funding from the European Union’s Horizon 2020 research and innovation program under the Marie Skodowska-Curie Grant Agreement No. 872081.
 This work is also supported by the Consejería de Transformación Económica, Industria, Conocimiento y Universidades, Junta de Andalucía and European Regional Development Fund (ERDF 2021-2027) under the project EPIT1462023.
J.K.-R. also acknowledges support from a
Spanish Ministerio de Universidades “Margarita Salas” Fellowship. Computing resources supporting this work were
provided by the CEAFMC and Universidad de Huelva High Performance Computer located in the Campus Universitario
“El Carmen” and funded by FEDER/MINECO Project No. UNHU-15CE-2848.
\end{acknowledgments}
\appendix

\section{Analytical exponent} \label{App:class}
Finding analytically the exponent $\delta$ which governs the decay to zero of the quasi-degenerate gap in the classical limit could be a laborious task. We know that due to the finite-high barrier, the tunneling probability between the two wells is not zero~\cite{messiah2020}. Also, we know that the tunneling effect decays exponentially with $N_e=1/\heff$~\cite{Landau1991}. On the basis of that events, we are going to use a semiclassical analysis to find the exponent $\delta$ in $\Delta_n\propto e^{-\delta N_e}$.

For simplicity, we will focus on the parity-symmetric case,
\begin{equation}\label{eq:kerrH_app}
    \hat{h}(\xi_2)=\frac{\hat{H}(\xi_2,\xi_1=0)}{K\hbar}=  \frac{\hat{a}^{\dagger 2}\hat{a}^2}{N_e}- \xi_2\left(\hat{a}^{\dagger 2} + \hat{a}^2\right)~.
\end{equation}
We can use a Glauber coherent state~\cite{CohSta1990}
\begin{equation}
    \ket{\zeta}=e^{-\frac{|\zeta|^2}{2}} \sum_{n=0}^{\infty} \frac{\zeta^n}{\sqrt{n!}}\ket{n}~,
\end{equation}
where $\zeta$ is a variational complex parameter linked with $(q,p)$ by the relation $\zeta=\frac{1}{\sqrt{2}}\left( q+ip\right)$. Now, we can define an energy functional as follows
\begin{align}
    h(\xi_2)&=\frac{|\zeta|^4}{N_e} - \xi_2 \left(\overline{\zeta}^2+\zeta^2\right)\\
        &= \frac{1}{4N_e}\left(p^2+q^2\right)^2 - \xi_2 (q^2-p^2)~. \nonumber
\end{align}
Minimizing the energy functional, we find that there exist five different extrema for $\xi_2\neq 0$, a maximum and four minima, which cannot exist simultaneously,
\begin{align}
    X_0&=\left(0,0\right) \nonumber\\
    X_{\pm}^q&=\left(\pm\sqrt{2\xi_2N_e},0\right) \\
    X_{\pm}^p&=\left(0,\pm\sqrt{-2\xi_2N_e}\right)~. \nonumber
\end{align}
Since our problem is now parity symmetric, we can restrict the $\xi_2$ values to either $\xi_2>0$ or $\xi_2<0$ without loss of generality. 

It is known that the basis of coherent states is not orthogonal, then the overlap between two different coherent states is given by~\cite{CohSta1990}
\begin{equation}
    \braket{\zeta}{\zeta'}=e^{\overline{\zeta}\zeta' - \frac{1}{2}|\zeta|^2 - \frac{1}{2}|\zeta'|^2}~.
\end{equation}
Now, we only need to place a coherent state in each well, $\ket{  \zeta_{\pm}=\pm\sqrt{|\xi_2|N_e}}$, and calculate the overlap between them
\begin{equation}
    \braket{\zeta_+=\sqrt{|\xi_2|N_e}}{\zeta_-= -\sqrt{|\xi_2|N_e}} = e^{-2|\xi_2|N_e}~.
\end{equation}
This result is in concordance with what we observed, $\delta=2|\xi_2|$.

\section{Dependence on the one-photon squeezing \label{app:QP}}
\begin{figure}[h]
    \centering
    \includegraphics[width=\columnwidth]{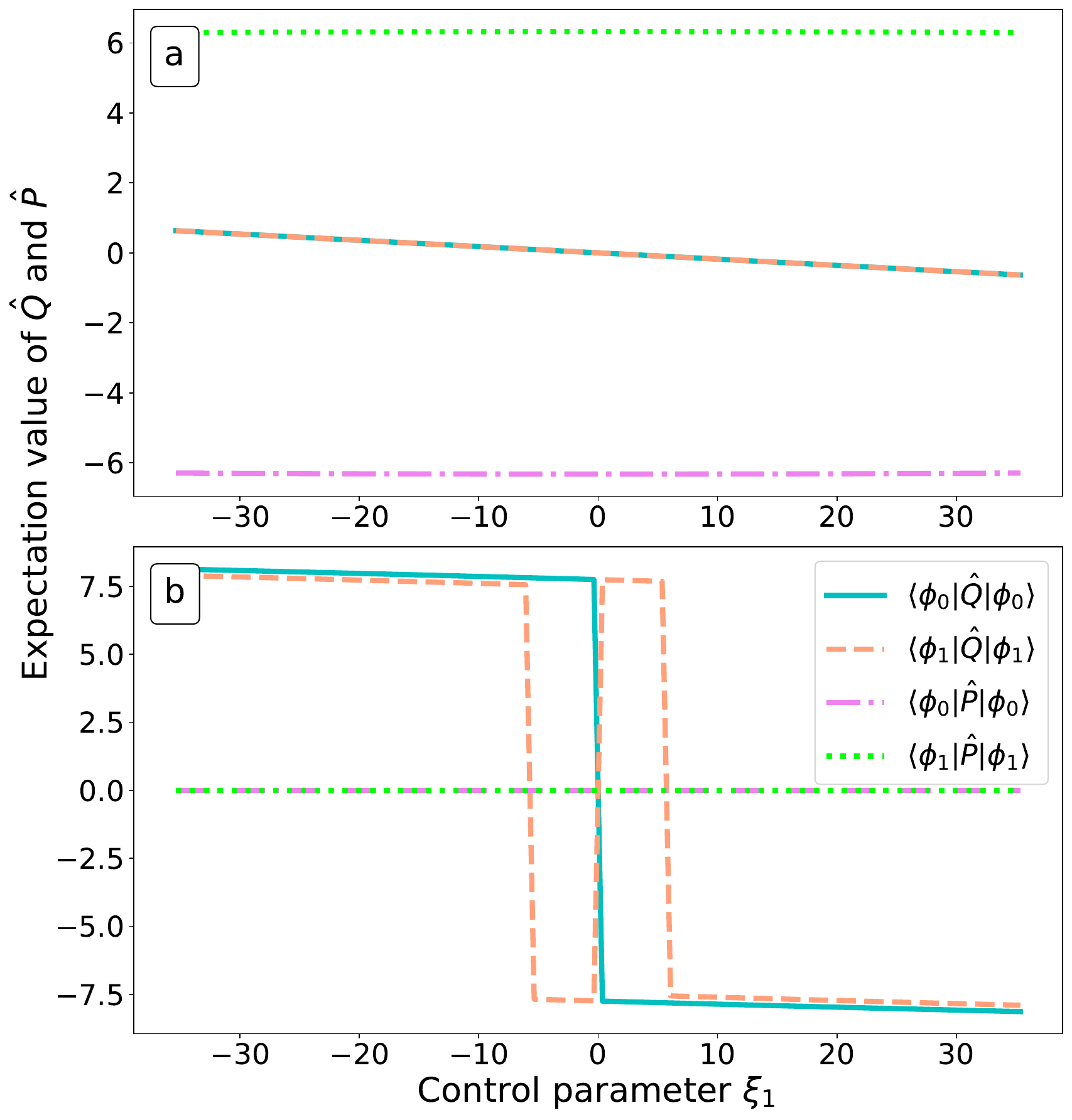}
    \caption{Panels (a) and (b) show the expectation value of the coordinate and momentum operators, $\hat Q$ and $\hat{P}$, for the two lowest energy eigenstates, $\ket{\phi_0}$ and $\ket{\phi_1}$, versus the control parameter $\xi_1$ for $\xi_2 = -30$ in panel (a) and $30$ in panel (b). Quantum calculations have been carried out with $N_e=1$.}
    \label{fig:Correlationxi1_app}
\end{figure}

A better understanding of the dependence of Hamiltonian~\eqref{eq:kerrH} on the one-photon squeezing can be achieved by studying the coordinate and momentum averages. The expectation values of the  $\hat{Q}$ and $\hat{P}$ operators confirm the results obtained and are in perfect agreement with the classical limit shown in Fig.~\ref{fig:Correlationxi1}. In the   $\xi_2 = -30$ case, shown in  Fig.~\ref{fig:Correlationxi1_app}(a), the $\hat{Q}$  expectation values are small and change from positive to negative values as $\xi_1$ goes through zero, while the $\hat{P}$ expected values are positive or negative, both results are in agreement with the energy surfaces shown in panels I and II of Fig.~\ref{fig:Correlationxi1}(a). In the  $\xi_2 = 30$ case, the results obtained are depicted in  Fig.~\ref{fig:Correlationxi1_app}(b), where the expectation values of  $\hat{Q}$ for the ground state, $\ket{\phi_0}$, switch sign when $\xi_1$ goes through zero. In this case, the obtained results for $\ket{\phi_1}$ are more complex, changing sign three times, once on the origin and twice for positive and negative $\xi_1$ values that can be linked to crossings in the energy level diagram. As regards the expected $\hat{P}$ values, they  remain equal to zero for the two states. The results shown in  Fig.~\ref{fig:Correlationxi1_app}, together with Fig.~\ref{fig:Correlationxi1}, allow for a better understanding of the differences between positive and negative $\xi_2$ ranges in Fig.~\ref{fig:Correlationxi2}, and clearly point towards the existence of a first order ground-state QPT in this system. This is something that we intend to explore in future works, as well as the hint of possible local symmetries associated with the location of the crossings in this case, something already found in the pure two-photon driven case~\cite{Iachello2023}.

\bibliography{refs.bib}
\bibliographystyle{apsrev4-2}

\end{document}